\documentclass[sigconf]{acmart}
\AtBeginDocument{%
  \providecommand\BibTeX{{%
    \normalfont B\kern-0.5em{\scshape i\kern-0.25em b}\kern-0.8em\TeX}}}

\setcopyright{acmcopyright}
\copyrightyear{2025}
\acmYear{2025}
\acmDOI{XXXXXXX.XXXXXXX}

\acmConference[Conference acronym 'XX]{Make sure to enter the correct
  conference title from your rights confirmation email}{June 03--05,
  2018}{Woodstock, NY}
\acmPrice{15.00}
\acmISBN{978-1-4503-XXXX-X/18/06}

\usepackage{cleveref}
\usepackage{subcaption}
\usepackage{enumitem}
\usepackage{algorithm}
\usepackage{algpseudocode}
\usepackage{amsmath}
\usepackage{bbm}
\begin{document}

\title{SurvEval: Listwise Re-Ranking Evaluator via Survival-Based Exposure and Multi-Objective Value Modeling}

\author{YaChen Yan}
\email{yachen_yan@intuit.com}
\orcid{0000-0002-1213-4343}
\affiliation{%
    \institution{Intuit Credit Karma}
    \streetaddress{760 Market Street}
    \city{Mountain View}
    \state{California}
    \country{USA}
    \postcode{94102}
}

\author{Liubo Li}
\email{liubo_li@intuit.com}
\orcid{1234-5678-9012}
\affiliation{%
    \institution{Intuit Credit Karma}
    \streetaddress{760 Market Street}
    \city{Mountain View}
    \state{California}
    \country{USA}
    \postcode{94102}
}

\renewcommand{\shortauthors}{YaChen and LiuBo}
\renewcommand{\subtitle}{SurvEval: Listwise Re-Ranking Evaluator via Survival-Based Exposure and Multi-Objective Value Modeling}


\begin{abstract}
Re-ranking is a crucial stage in recommender systems, yet existing evaluators oversimplify user browsing assumptions and collapse multi-objective signals into a single scalar, limiting their ability to reflect true listwise utility. We propose \textbf{SurvEval}, a listwise evaluator that jointly models user exposure and item value for re-ranking evaluation. We formalize the sequential browsing process as a discrete-time survival model, deriving monotonically decreasing reaching probabilities that capture the cascading structure of user browsing. We further introduce a multi-objective value estimation module with dedicated per-objective prediction heads and uncertainty-based loss weighting to adaptively balance objectives of varying scales and noise levels. Both components are jointly optimized via a unified transformer backbone, and the listwise utility is obtained by summing the position-wise products of reaching probabilities and conditional item value estimates. Extensive experiments on KuaiRand and Avito demonstrate that SurvEval achieves superior performance over state-of-the-art baselines.
\end{abstract}



\maketitle

\section{Introduction}
Recommender systems operate through a cascade of stages: retrieval narrows a large corpus to a candidate set, a ranking stage refines scores, and a re-ranking stage optimizes the final presented list~\cite{covington2016deep}. The re-ranking stage must account not only for individual item relevance, but also for inter-item interactions and the sequential nature of user browsing.
 
A reliable evaluator must capture two fundamental aspects: \emph{how far} a user browses into the list, and \emph{how much value} the user derives from observed items. These two factors are tightly coupled: an item that is never reached contributes nothing to utility regardless of its intrinsic quality, making listwise utility estimation inherently non-trivial.
 
Existing evaluators fall short on both dimensions. On the exposure side, most methods model list context without explicitly accounting for the probability that a user actually reaches each position~\cite{ai2018learning, pei2019personalized}, ignoring the cascading, user-dependent nature of browsing. In reality, the probability of observing an item at position $t$ depends on the user's decision to continue past all preceding positions. On the value side, existing approaches aggregate multiple engagement signals (clicks, watch time, likes) into a fixed scalar combination~\cite{shi2023pier, yang2025comprehensive}, conflating objectives of different scales and noise levels.

We propose \textbf{SurvEval}, a listwise evaluator that decomposes list utility into position-wise contributions by jointly modeling exposure and value. Central to our approach is the observation that a user can only derive value from an item they actually see: the expected utility of a ranked list is therefore naturally decomposed as a sum of per-position terms, each being the product of the \emph{reaching probability} (the probability that the user browses as far as position $t$) and the item's \emph{conditional value} given observation. This decomposition motivates two dedicated modeling components. For exposure, we formalize sequential browsing as a discrete-time survival process, yielding monotonically decreasing reaching probabilities without hand-crafted discount functions. For value, we adopt a multi-objective framework with per-objective prediction heads and uncertainty-based multi-task weighting. Both components share a unified transformer backbone.

\section{Related Work}
Reranking models broadly fall into one-stage and two-stage methods.

One-stage methods model the target ranking in a single framework. Early works such as DLCM and PRM incorporate list context into deep models to refine item scores~\cite{ai2018learning, pei2019personalized}, while SetRank uses permutation-invariant representations for set-level modeling~\cite{pang2020setrank}. Other approaches treat rankings as structured outputs emphasizing global optimization~\cite{zhuang2018globally}, and more recent generative methods produce ranked lists directly~\cite{bello2018seq2slate, feng2021grn, ren2024non, wang2025nlgr, zhang2025goalrank, lin2025gref}.

Two-stage methods decompose re-ranking into a generation stage and an evaluation stage. Representative work includes PIER, which combines permutation-level generation with interest-aware evaluation end-to-end~\cite{shi2023pier}, and subsequent approaches that improve generation diversity through multi-generator construction~\cite{yang2025comprehensive}. More recently, evaluator-focused work has explored list-level value estimation to better align re-ranking with user utility~\cite{zhang2025generation}, which directly motivates our work.

\section{Methodology}

\subsection{Framework Overview}

Given a user and a candidate list from an upstream ranker, our listwise evaluator assigns a scalar utility to the entire list by jointly modeling per-position user exposure and item value. The evaluator is trained end-to-end on logged recommendation data with observed exposure and engagement signals.

\subsection{Problem Formulation}

We consider a multi-stage recommendation setting with a user space $\mathcal{U}$ and an item space $\mathcal{I}$. Let $u \in \mathcal{U}$ denote a user and $I \in \mathcal{I}$ denote an item. Given a user $u$, an upstream ranking stage produces a candidate set of $m$ items. The re-ranking task selects and orders $n \leq m$ items into a final displayed list $L = (I_{l_1}, I_{l_2}, \dots, I_{l_n})$, where $n \ll m$. The goal of listwise evaluation is to assign a scalar utility $r(u, L)$ to the ordered list $L$ for user $u$, reflecting the expected total engagement the list would generate under realistic user browsing behavior. A faithful evaluator must account for two interacting factors: how far the user browses into the list, and how much value each item generates upon being seen.

\subsection{User and Item Encoders}

A key component of our framework is a shared representation scheme that maps users and items into dense embedding tokens for downstream modeling. We introduce a user encoder and an item encoder to transform each user and item into dense representations, referred to as user tokens and item tokens respectively.

\subsubsection{User Encoder}
The user encoder produces a fixed set of user tokens from three components: a linear projection of static features split into multiple tokens; a multi-interest extraction layer~\cite{cen2020controllable} that aggregates long-term user interactions into tokens via softmax attention; and the most recent $N$ user interactions as tokens. Outputs are concatenated and RMS-normalized.

\begin{figure}[htpb]
    \centering
    \includegraphics[width=0.40\textwidth, height=0.20\textwidth]{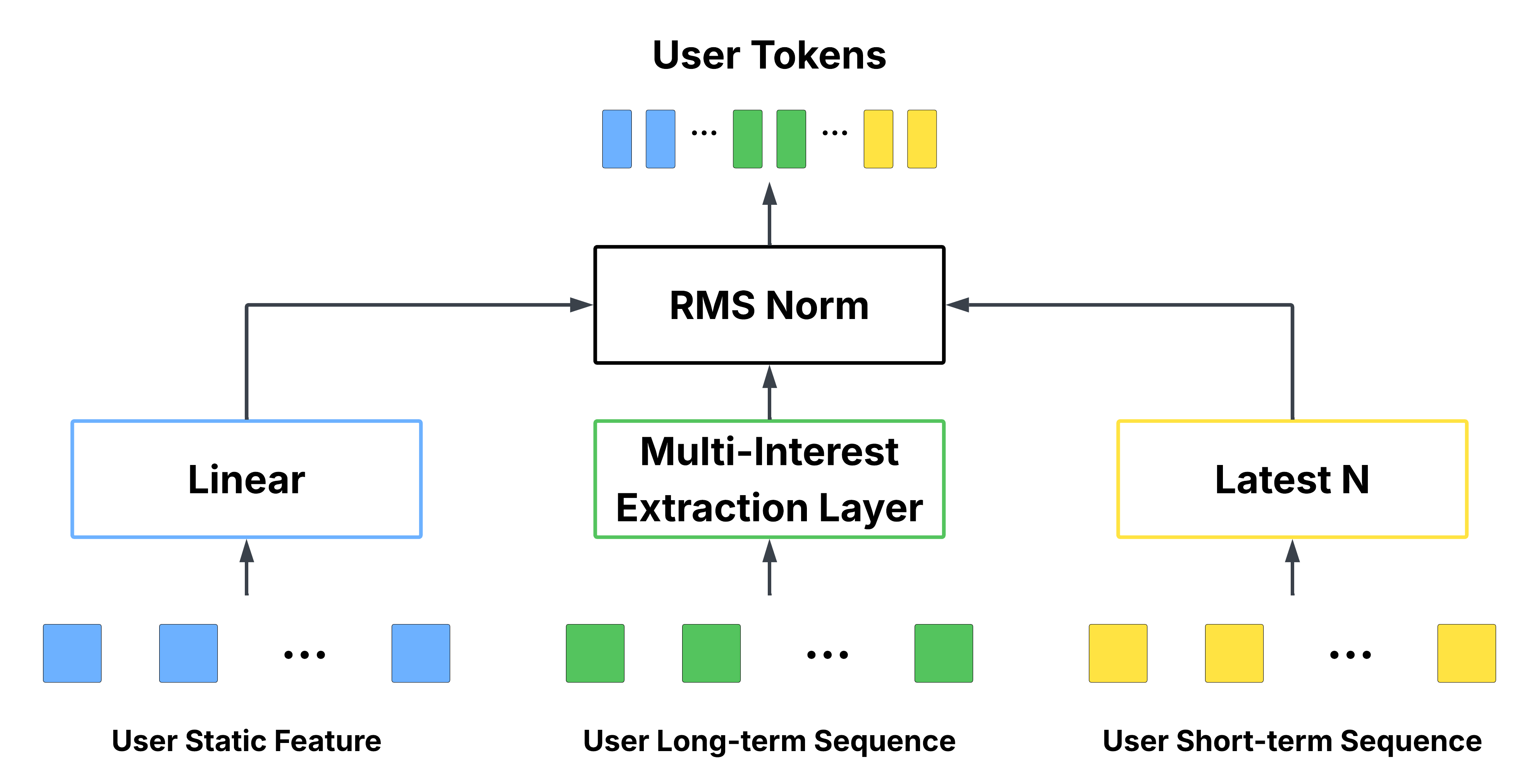}
    \caption{The user encoder maps heterogeneous user signals into a fixed set of user tokens, which are normalized to form a unified representation for downstream modules.}
    \label{fig_user_encoder}
    \medskip
    \small
\end{figure}

\subsubsection{Item Encoder}
Each item is encoded by element-wise adding a trainable item ID embedding and a linearly projected side-information representation (e.g., category, price), followed by RMS normalization.

\subsection{Listwise Utility Decomposition}

We introduce a listwise evaluator that assigns a scalar utility to an ordered list for a given user. Given a user $u$ and an ordered list $L = (I_{l_1}, \dots, I_{l_n})$, the listwise utility is defined as 
$r_\phi(u, L)$, where $\phi$ denotes the model parameters.

Rather than directly modeling the utility of the entire list, we decompose it into position-wise contributions under the user browsing process:
\begin{equation}
r_\phi(u, L) = \sum_{t=1}^{n} 
\underbrace{p_\phi(t \mid u, L)}_{\text{reaching probability}} 
\cdot \underbrace{v_\phi(u, I_{l_t})}_{\text{item value}},
\end{equation}
where $p_\phi(t \mid u, L)$ denotes the probability that the user reaches and observes the item at position $t$, and $v_\phi(u, I_{l_t})$ denotes the expected value of the item conditional on being observed. This formulation captures the intuition that each item contributes to list utility proportionally to its observation probability, naturally motivating the joint modeling of exposure and value.

\subsection{Evaluator Architecture}

\begin{figure}[htpb]
    \centering
    \includegraphics[width=0.40\textwidth, height=0.40\textwidth]{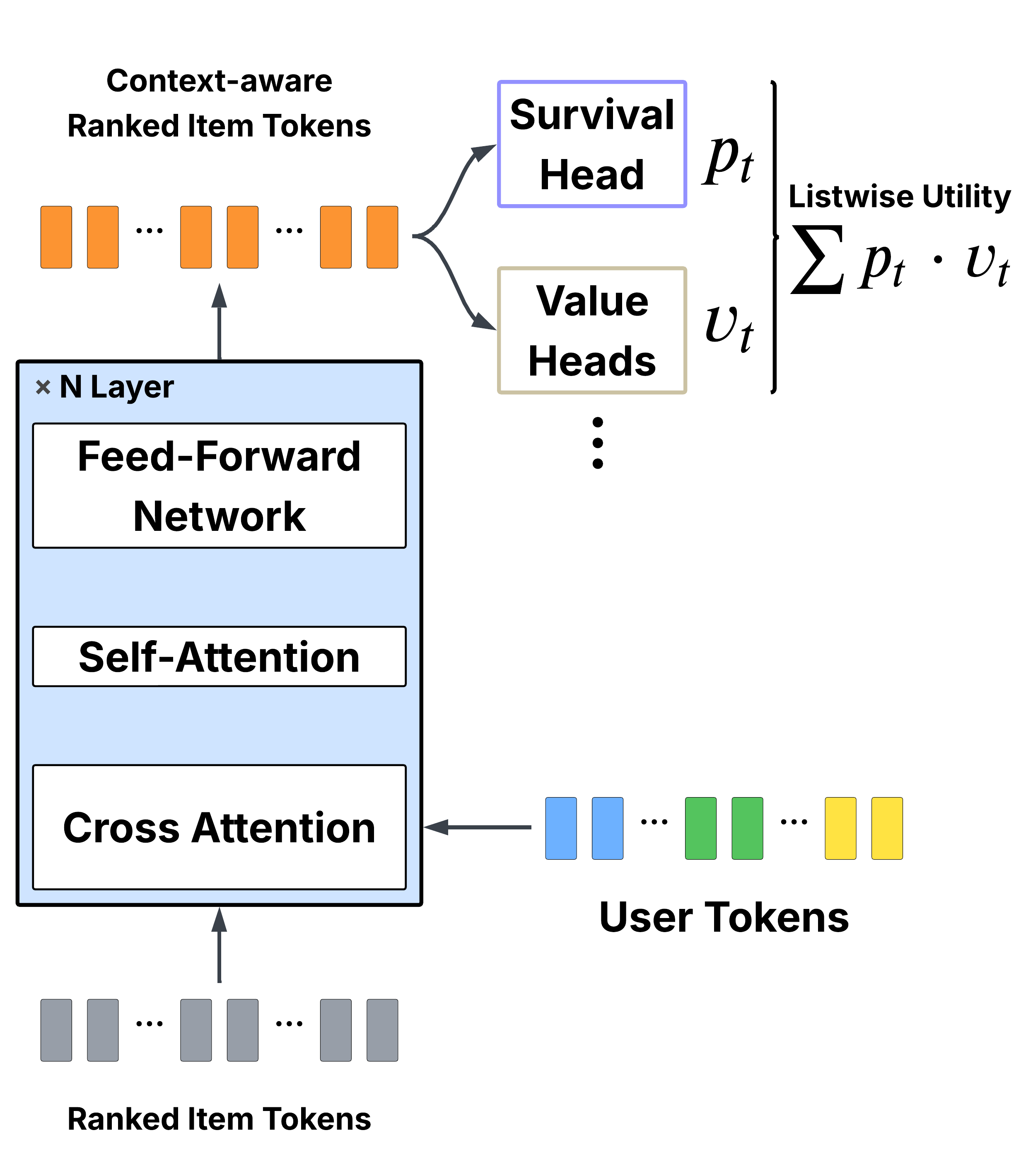}
    \caption{Architecture of the listwise evaluator. A transformer backbone encodes ranked item tokens via cross-attention to user tokens and self-attention over the ranked sequence, producing context-aware representations fed into a survival head and $K$ value heads. The listwise utility is computed as $\sum_t p_\phi(t) \cdot v_t$, where $v_t = \sum_k w_k \hat{v}_t^{(k)}$.}
    \label{fig_evaluator}
    \medskip
    \small
\end{figure}

To jointly model reaching probability and item value, we employ a unified transformer-based architecture over the ordered list, as illustrated in \autoref{fig_evaluator}. Each ranked item token is first summed with a learned position embedding, as position directly governs exposure and contextual perception. The position-augmented tokens are then processed by a stack of transformer layers.

The resulting context-aware ranked item tokens are then fed into two types of task-specific heads:
\begin{itemize}
    \item a \textbf{survival head} that predicts the reaching probabilities,
    \item $K$ \textbf{value heads}, for per-objective value estimation.
\end{itemize}

\subsection{Exposure Modeling}

We model $p_\phi(t \mid u, L)$ as the probability that the user reaches and observes the item at position $t$, where $E_t \in \{0,1\}$ denotes whether position $t$ is observed.

\subsubsection{Sequential Browsing Intuition}
Users scroll top-down and may exit at any position, making exposure inherently cascading: reaching position $t$ requires surviving all preceding positions, a structure naturally captured by a discrete-time survival process.

\subsubsection{Survival-Based Formulation}
We formalize this as a discrete-time survival process over $n$ positions. Let $M$ denote the exit position. The fundamental quantity is the \textbf{conditional hazard}:
\begin{equation}
    h_\phi(t \mid u, L) = P(M = t \mid M \geq t,\ u, L),
\end{equation}
the probability of exit at position $t$ given the user has reached it. The reaching probability is then derived as:
\begin{equation}
    p_\phi(t \mid u, L) = \prod_{j=1}^{t-1} \bigl(1 - h_\phi(j \mid u, L)\bigr),
\end{equation}
with $p_\phi(1 \mid u, L) = 1$ by convention. At each position, the user exits with probability $h_\phi(t)$ or continues with probability $1 - h_\phi(t)$, yielding a monotonically decreasing reaching probability with depth.

\subsubsection{Exposure Modeling Objective}
Let $C = \min(M, n)$ denote the censoring position, and $\delta = \mathbf{1}[M \le n]$ the event indicator. The likelihood of an observed browsing trajectory under the discrete-time survival process is $P(C, \delta \mid u, L) = h_\phi(C \mid u, L)^{\delta} \prod_{t=1}^{C-\delta} \bigl(1 - h_\phi(t \mid u, L)\bigr)$, yielding the negative log-likelihood
\begin{equation}
\mathcal{L}_{\text{exp}} = -\,\delta \log h_\phi(C \mid u, L) \;-\; \sum_{t=1}^{C-\delta} \log\bigl(1 - h_\phi(t \mid u, L)\bigr).
\end{equation}
Equivalently, defining per-position exit indicators $y_t = \delta\,\mathbf{1}[t = C]$ over the reached positions $t \le C$, $\mathcal{L}_{\text{exp}}$ is the binary cross-entropy between $y_t$ and $h_\phi(t \mid u, L)$; right-censored trajectories ($\delta = 0$) contribute survival terms at every reached position and no event term.

\begin{figure}[htpb]
    \centering
    \includegraphics[width=0.30\textwidth, height=0.25\textwidth]{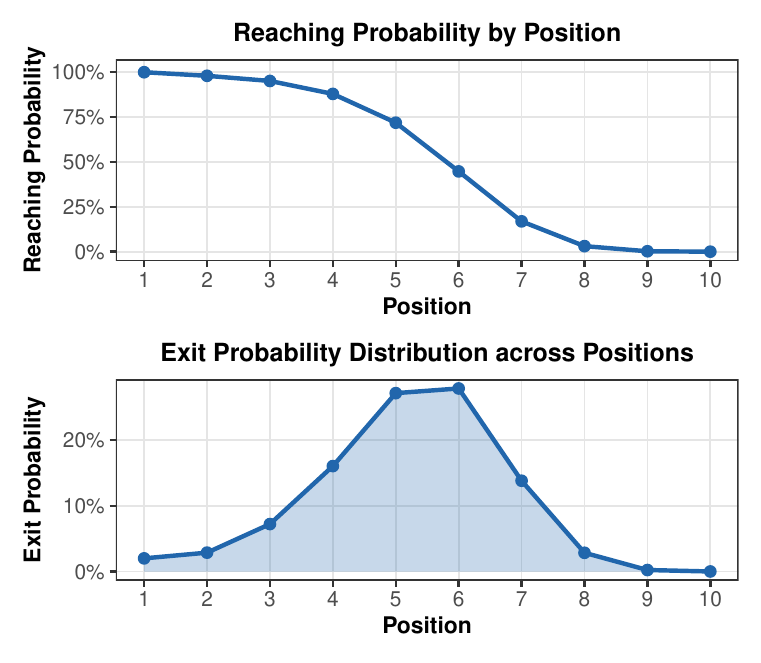}
    \caption{Illustrative reaching and exit probability profiles under the survival-based browsing model. \textbf{Top:} Reaching probability decreases monotonically with position depth. \textbf{Bottom:} Exit probability peaks at intermediate positions, following a unimodal distribution.}
    \label{fig_survival}
\end{figure}

\subsection{Value Modeling}
Rather than collapsing multiple engagement objectives (clicks, watch time, likes, shares) into a single scalar, we model each independently with a dedicated prediction head. Formally, let $\mathcal{O} = \{1, \dots, K\}$ be the set of objectives, and $Y_t^{(k)} \geq 0$ the sparse, non-negative observed utility for objective $k$ at position $t$. The aggregated item value is a weighted sum of per-objective estimates:


\begin{equation}
v_\phi(u, I_{l_t}) = \sum_{k=1}^{K} w_k \cdot \hat{v}_\phi^{(k)}(u, I_{l_t}),
\end{equation}
where $\hat{v}_\phi^{(k)}(u, I_{l_t})$ is the per-objective value estimate for objective $k$, and $w_k$ is the manually specified importance weight.

\subsubsection{Per-Objective Head and Weighted Logistic Regression}
Each objective $k$ is modeled via a dedicated value head that produces a scalar logit $f_\phi^{(k)}(u, I_{l_t})$ from the shared context-aware representation. Following~\citet{covington2016deep}, we adopt weighted logistic regression to handle sparse, non-negative utility labels. Positive examples (i.e., $Y_t^{(k)} > 0$) are weighted by their observed utility value $Y_t^{(k)}$, while negative examples (i.e., $Y_t^{(k)} = 0$) receive unit weight. Under this scheme, the learned odds closely approximate the expected utility $\mathbb{E}[Y^{(k)}]$. At inference time, the exponential function recovers these odds as the per-objective value estimate:
\begin{equation}
\hat{v}_\phi^{(k)}(u, I_{l_t}) = \exp\!\left(f_\phi^{(k)}(u, I_{l_t})\right).
\end{equation}

\subsubsection{Per-Objective Training Loss}
The per-objective loss is a weighted cross-entropy over observed positions:
\begin{equation}
\mathcal{L}_{\text{value}}^{(k)} =
-\sum_{t:\, E_t = 1}
\left[
Y_t^{(k)} \cdot \log \sigma\!\left(f_\phi^{(k)}(u, I_{l_t})\right)
+ \log \left(1 - \sigma\!\left(f_\phi^{(k)}(u, I_{l_t})\right)\right)
\right],
\end{equation}
where $\sigma(\cdot)$ is the sigmoid function. Positive examples with larger $Y_t^{(k)}$ contribute proportionally more to the loss, guiding the model to accurately predict the magnitude of the utility rather than merely its occurrence.

\subsubsection{Uncertainty-Based Multi-Task Weighting}
A key challenge in multi-objective learning is balancing losses across objectives with different scales and noise levels. Following~\citet{kendall2018multi}, we adopt an uncertainty-based multi-task learning approach that automatically learns per-task loss weights by modeling the homoscedastic uncertainty of each objective. For each objective $k$, we introduce a learnable log-uncertainty parameter $s_k = \log \sigma_k^2$, where $\sigma_k > 0$ captures the observation noise of task $k$. The uncertainty-weighted loss for objective $k$ is: $\widetilde{\mathcal{L}}_{\text{value}}^{(k)} = e^{-s_k} \mathcal{L}_{\text{value}}^{(k)} + \frac{1}{2} s_k$.

The term $e^{-s_k}$ down-weights objectives with higher uncertainty, while the regularization term $\frac{1}{2} s_k$ prevents the model from trivially driving $\sigma_k \to \infty$. The parameters $\{s_k\}_{k=1}^{K}$ are optimized jointly with the model parameters $\phi$.

\subsubsection{Value Modeling Objective}
The overall value training objective is:
\begin{equation}
\mathcal{L}_{\text{value}} = \sum_{k=1}^{K} \widetilde{\mathcal{L}}_{\text{value}}^{(k)}
= \sum_{k=1}^{K} \left( e^{-s_k} \mathcal{L}_{\text{value}}^{(k)} + \frac{1}{2} s_k \right).
\end{equation}

\subsection{Training and Inference}

\subsubsection{Training}
The evaluator models a joint distribution over exposure and value:
$P(E_{1:n}, L_{1:n}) = P(E_{1:n}) \cdot P(L_{1:n} \mid E_{1:n})$, and is trained by minimizing:
\begin{equation}
\mathcal{L} = \mathcal{L}_{\text{exp}} + \mathcal{L}_{\text{value}}.
\end{equation}

where $\mathcal{L}_{\text{exp}}$ is learned from user browsing trajectories and $\mathcal{L}_{\text{value}}$ from item-level engagement signals on exposed positions.

\subsubsection{Inference}
The listwise utility is obtained by multiplying each item's aggregated value estimate by its reaching probability, then summing across all positions:
\begin{equation}
    r_\phi(u, L) = \sum_{t=1}^{n} p_\phi(t \mid u, L) \cdot v_\phi(u, I_{l_t}),
\end{equation}
where $v_\phi(u, I_{l_t}) = \sum_{k=1}^{K} w_k \cdot \hat{v}_\phi^{(k)}(u, I_{l_t})$ is the aggregated item value. Items at deeper positions are discounted by cumulative exit probability, incentivizing high-value items to be placed where hazard is low.

At serving time, following the Multi-Generator Evaluator (MG-E) framework~\cite{yang2025comprehensive}, multiple generators produce candidate lists, and the evaluator selects the list with the highest utility $r_\phi(u, L)$ for final exposure.





\section{Experiments}

\subsection{Experimental Settings}

\subsubsection{Datasets}
\textbf{KuaiRand}\footnote{\url{https://kuairand.com/}} is a large-scale randomly logged dataset from Kuaishou with multi-objective engagement signals (watch time, likes, follows, comments). Random item insertion provides unbiased exposure signals suited for exposure-aware evaluation. The exit position is inferred as the first position after which all subsequent play times fall below a minimum threshold. \textbf{Avito}\footnote{\url{https://www.kaggle.com/c/avito-context-ad-clicks}} contains user search page logs with binary ad click signals. Exit is inferred at the position of the last clicked ad.

\subsubsection{Evaluation Metrics}
We report \textbf{NDCG}, which rewards relevant items at higher positions with logarithmic discounting, and \textbf{uAUC}, a user-level generalized AUC averaged across users that supports non-binary labels and is robust to user activity imbalance.

\subsubsection{Baselines}
We compare against: pointwise model \textbf{DNN}~\cite{covington2016deep}; listwise contextual model \textbf{PRM}~\cite{pei2019personalized}; and permutation-level model \textbf{PIER}~\cite{shi2023pier} and \textbf{CAVE}~\cite{zhang2025generation}.

\subsubsection{Implementation Details}
All models are in PyTorch. User encoder: 8 long-term interest tokens, $N{=}20$ short-term tokens. Item encoder: 64-dim ID embedding plus projected side information. Transformer: 2 layers, 4 heads, hidden dimension 128, learned position embeddings. Objectives: $K{=}1$ (Avito), $K{=}4$ (KuaiRand), uniform $\{w_k\}$, $\{s_k\}$ initialized to 0. Optimizer: Adam, lr $1{\times}10^{-4}$, batch size 256, early stopping on validation NDCG (patience 3).

\subsection{Overall Performance}
As shown in \autoref{tab_main}, SurvEval consistently achieves the best results across both datasets and metrics. List-level models outperform the pointwise DNN baseline, and SurvEval further improves over all baselines, validating the survival-based exposure model and disentangled multi-objective value estimation.

\begin{table}[htpb]
\caption{Overall performance comparison on KuaiRand and Avito datasets.}
\label{tab_main}
\begin{tabular}{lcccc}
\toprule
  & \multicolumn{2}{c}{KuaiRand} 
  & \multicolumn{2}{c}{Avito} \\
\cmidrule(lr){2-3} \cmidrule(lr){4-5}
  & NDCG & uAUC
  & NDCG & uAUC \\
\midrule
DNN         & 0.7965 & 0.6781 & 0.7022 & 0.6956 \\
PRM         & 0.8180 & 0.6973 & 0.7182 & 0.7019 \\
PIER        & 0.8312 & 0.7091 & 0.7281 & 0.7183 \\
CAVE        & 0.8450 & 0.7196 & 0.7371 & 0.7258 \\
\midrule
SurvEval    & \textbf{0.8523} & \textbf{0.7236} & \textbf{0.7450} & \textbf{0.7326} \\
\bottomrule
\end{tabular}
\end{table}

\section{Conclusion}

We presented SurvEval, a listwise re-ranking evaluator that decomposes list utility into position-wise contributions by jointly modeling survival-based user exposure and multi-objective item value through a unified transformer architecture. Experiments on KuaiRand and Avito show consistent improvements over state-of-the-art baseline methods. Future work includes extending the survival model to non-sequential browsing patterns, learning importance weights in a user-adaptive manner, and integrating the evaluator with a jointly trained generator.

\newpage
\bibliographystyle{ACM-Reference-Format}
\bibliography{surveval.bib}

\end{document}